\documentclass[article]{elsarticle}

\usepackage{lineno,hyperref}
\modulolinenumbers[5]

\journal{Journal of Parallel and Distributed Computing}
\usepackage{graphicx, subfigure}
\usepackage{framed}
\usepackage[flushleft]{threeparttable}
\usepackage{multirow}
\usepackage{hhline}
\usepackage{lipsum}
\usepackage{algpseudocode}
\usepackage{algorithm}
\usepackage{soul, color}
\usepackage{hyperref}
\soulregister\cite7
\soulregister\url7
\soulregister\ref7

\algnewcommand{\LineComment}[1]{\Comment #1}
\algrenewcommand\textproc{}

\renewcommand{\hl}[1]{#1}

\begin{document}

\begin{frontmatter}
\title{LightScan: Faster Scan Primitive on CUDA Compatible Manycore Processors}

\author{Yongchao Liu
}
\ead{yliu@cc.gatech.edu}
\author{Srinivas Aluru
}
\ead{aluru@cc.gatech.edu}

\address{School of Computational Science \& Engineering\\
 Georgia Institute of Technology\\Atlanta, USA}

\begin{abstract}
Scan (or prefix sum) is a fundamental and widely used primitive in parallel computing. In this paper, we present LightScan, a faster parallel scan primitive for CUDA-enabled GPUs, which investigates a hybrid model combining intra-block computation and inter-block communication to perform a scan. Our algorithm employs warp shuffle functions to implement fast intra-block computation and takes advantage of globally coherent L2 cache and the associated parallel thread execution (PTX) assembly instructions to realize lightweight inter-block communication. Performance evaluation using a single Tesla K40c GPU shows that LightScan outperforms existing GPU algorithms and implementations, and yields a speedup of up to $2.1$, $2.4$, $1.5$ and $1.2$ over the leading CUDPP, Thrust, ModernGPU and CUB implementations running on the same GPU, respectively. Furthermore, LightScan runs up to $8.9$ and $257.3$ times faster than Intel TBB running on 16 CPU cores and an Intel Xeon Phi 5110P coprocessor, respectively. \hl{Source code of LightScan is available at \url{http://cupbb.sourceforge.net}.}
\end{abstract}

\end{frontmatter}

\section{Introduction}
As an important operation in parallel computing, scan (or prefix sum) primitive \cite{blelloch1989scans} is frequently used in various applications such as sequence alignment \cite{aluru2003parallel} \cite{khajeh2010acceleration}, graph algorithms \cite{vineet2009fast}~\cite{merrill2012scalable}, parallel sort \cite{cederman2009sorting}~\cite{satish2009designing}, cloud computing \cite{fang2011mars}, sparse linear algebra \cite{zhang2012novel}, machine learning \cite{kohlhoff2013k}~\cite{niu2014fast} and suffix array construction \cite{liuparallel}~\cite{patrick2015}. There are two variants for scan primitive, namely inclusive scan and exclusive scan. Given a binary associative operator $\oplus$ and an array of $N$ elements {$x=\{x_0, x_1, \ldots, x_{N-2}, x_{N-1}\}$, an inclusive scan on $x$ computes a new array $y$ of size $N$ elements, with the $j$-th (\mbox{$0\leq j < N$}) element \mbox{$y_j = \bigoplus_{i=0}^j x_i$}. In other words, $y_j$ is the reduction of all elements in the prefix ending at the $j$-th position in $x$. Compared to inclusive scan, exclusive scan computes $y_j$ as \mbox{$\bigoplus_{i=0}^{j-1}x_i$}, i.e. the prefix reduction of all elements preceding the $j$-th element in $x$. Note that inclusive scan and exclusive scan can be mutually generated from each other. In this paper, we only focus on inclusive scan and also refer to it as simply as \textit{scan} unless otherwise specified. 
%

Graphics processing units (GPUs) are throughput-oriented manycore processors and have already become one of the most popular accelerators in high performance computing, particularly compute unified device architecture (CUDA)-based GPUs. A few GPU-based scan algorithms \cite{sengupta2007scan} \cite{harris2007parallel} \cite{dotsenko2008fast} \cite{wei2012optimization} \cite{yan2013streamscan} have been proposed in the literature. Typically, these algorithms work as follows. Firstly, the input array $x$ is decomposed into a set of consecutive and non-overlapping data blocks and one thread block, or work group in the context of open computing language (OpenCL)~\cite{stone2010opencl}, is assigned to locally scan one data block. Secondly, the reduction values for all blocks are first stored in an intermediate buffer and then scanned directly via inter-block computation \cite{harris2007parallel} \cite{dotsenko2008fast} or indirectly via inter-block communication \cite{yan2013streamscan}. Finally, each thread block updates its local scan accordingly with the help of the intermediate buffer, fulfilling the overall scan.

In this paper, we present LightScan, a faster parallel scan algorithm targeting CUDA-enabled GPUs with compute capabilities 3.0 or higher (currently only Kepler \cite{kepler} and Maxwell \cite{maxwell} GPU architectures). Similar to existing work in the literature, LightScan also partitions $x$ into a set of data blocks and assigns a thread block to process one block at a time. However, our algorithm introduces the following technical features/contributions. It ($i$) implements coalesced global loads and stores for $x$ and $y$ respectively, with exactly $N$ memory accesses each; ($ii$) saturates the use of registers per streaming multiprocessor (SM) to enlarge working set per thread block, and thereby reduces the number of inter-block communications; ($iii$) cyclically distributes data blocks over a fixed number of thread blocks in a deterministic manner, evading atomic-operation-based thread block re-indexing problem encountered in \cite{yan2013streamscan}; ($iv$) employs warp shuffle functions to implement fast intra-block local scan computation; ($v$) takes advantage of global coherence at L2 cache level (refer to the \textit{Cache Operators} section in \cite{ptxisa43}) and the associated parallel thread execution (PTX) assembly instructions to realize lightweight inter-block communication.

The performance of LightScan is evaluated using a set of $N$ values associated with four primitive scalar data types, i.e. 32-bit integer (Int32), 64-bit integer (Int64), single-precision floating-point (Float) and double precision floating-point (Double), on one Tesla K40c GPU with compute capability 3.5. We compare LightScan to several leading GPU-based algorithms, and Intel Threading Building Blocks (TBB) \cite{tbb} based implementations. The GPU-based algorithms tested include CUDPP \cite{harris2007parallel}, NVIDIA Thrust \cite{thrust}, NVIDIA ModernGPU \cite{moderngpu} and NVIDIA CUB \cite{cub}. These GPU-based implementations are all assessed on Tesla K40c, while TBB is separately evaluated on 16 CPU cores and an Intel Xeon Phi 5110P coprocessor. Performance evaluation shows that LightScan yields superior performance to the four GPU-based algorithms on the same GPU, as well as over Intel TBB either on 16 CPU cores or the Xeon Phi. More specifically, the speedup is on average $2.0$, $2.1$, $1.5$, $1.2$, $8.4$ and $80.8$ over CUDPP, Thrust, ModernGPU, CUB, TBB on 16 CPU cores, and TBB on the Xeon Phi, respectively.
\section{Related Work}
\label{sec:related_work}
Existing GPU-based scan algorithms consistently adopted a parallelization model, which first decomposes the input array $x$ into a set of data blocks and then employs a combination of inter-block and intra-block computation to complete the scan. For inter-block computation, two methods have been proposed: recursion (e.g. \cite{dotsenko2008fast}) and chaining (e.g. \cite{yan2013streamscan}). The recursion method recursively invokes scan operations on intermediate buffers, while the chaining method relies on direct communication of prefix reductions between neighbor blocks due to their serial dependency. Moreover, the recursion method needs frequent and relatively expensive global synchronization between thread blocks, while the chaining method requires relatively heavyweight peer-to-peer communications between thread blocks.

For intra-block computation, three parallel scan methods have been proposed, i.e. the Hillis-Steele scan \cite{hillis1986data}, work-efficient scan \cite{harris2007parallel}, and matrix-based scan \cite{dotsenko2008fast}, which will be briefly discussed in this section. For the convenience of discussion, we assume that there are $n$ elements per data block and $n$ threads to process a data block with each thread holding one element.
\subsection{Hillis-Steele Scan}
\label{sec:hs_scan}
The Hillis-Steele scan was proposed in \cite{hillis1986data} and then employed for GPU computing in \cite{horn2005stream}. Fig.~\ref{fig:hs_scan} shows the computational pattern and the pseodocode of the Hillis-Steele method. Given an input array $x$, this Hillis-Steele method perform \mbox{$\sum_{k=1}^{\log_2n}n-2^k$} = \mbox{$n(\log_2n-1)$} $\oplus$ operations. As each $\oplus$ requires two data loads and one data store, the whole computation will result in \mbox{$2n(\log_2n-1)$} loads and \mbox{$n(\log_2n-1)$} stores. If $x$ was deployed in shared memory or global memory, large number of memory accesses would result in bad performance.
\newcommand{\hsscankernel}{
\begin{algorithmic}[1]
\fontsize{8pt}{8.05pt}\selectfont
\Procedure{Hillis\_Steele\_scan}{$x$, $n$, $\oplus$}
    \For{$k$=0; $k <\log_2n$; ++$k$}
    	\ForAll{$0\leq i < n$ in parallel}
        	\If{$i\ge 2^k$}
            	\State $x_i$ = $x_i \oplus x_{i-2^k}$;
            \EndIf
        \EndFor
    \EndFor
\EndProcedure
\end{algorithmic}}

\begin{figure}[!h]
\centering
\begin{minipage}[c]{0.3\linewidth}
\centering
\includegraphics[width=\linewidth]{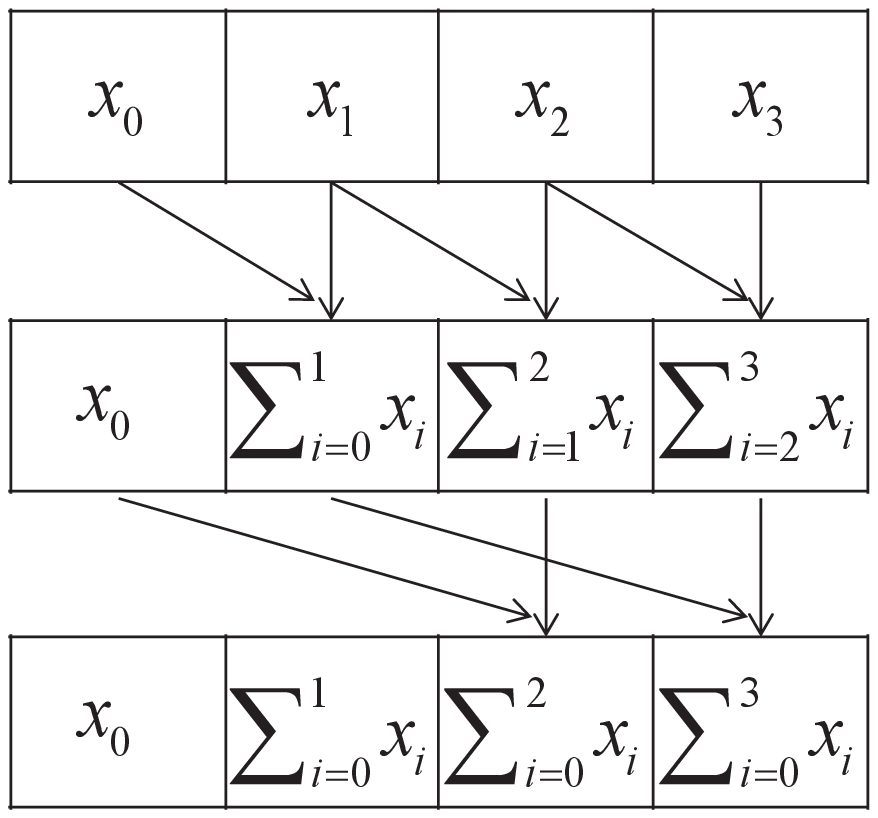}
\end{minipage}
\begin{minipage}[c]{0.68\linewidth}
\centering
\hsscankernel
\end{minipage}
\caption{Computational pattern and pseudocode of the Hillis-Steele parallel scan}
\label{fig:hs_scan}
\end{figure}
\subsection{Work-efficient Scan}
\label{sec:work_efficient}
To improve performance over the Hillis-Steele method, Harris \textit{et al.} \cite{harris2007parallel} proposed to use a work-efficient method consisting of two phases: up-sweep reduction and down-sweep accumulation, where the computation of each phase follows a complete binary tree topology. Fig.~\ref{fig:work_efficient} shows the computational patterns of the up-sweep and down-sweep phases. This work-efficient method performs \mbox{$2(n-1)$} $\oplus$ operations and \mbox{$n-1$} swap operations. Considering that each $\oplus$ requires two loads and one store and each swap two loads and two stores, it will require \mbox{$6(n-1)$} loads and \mbox{$4(n-1)$} stores to complete the scan, but significantly fewer than the Hillis-Steele method.
\begin{figure}[!h]
\centering
\includegraphics[width=0.55\linewidth]{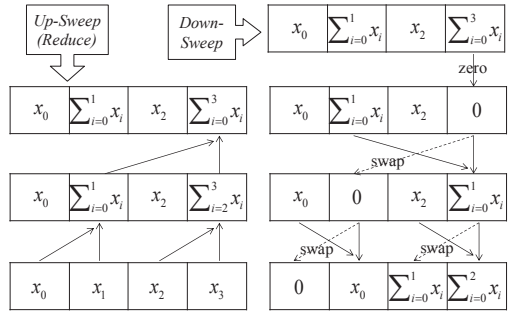}
\caption{Computational pattern of the work-efficient parallel scan}
\label{fig:work_efficient}
\end{figure}
\subsection{Matrix-based Scan}
\label{sec:matrix_scan}
Albeit efficient, the work-efficient method still has some performance bottlenecks such as ($i$) uncoalesced and long-latency accesses if global memory is used in the two phases; ($ii$) considerable bank conflicts if shared memory is used; and ($iii$) level-by-level synchronization for all threads within a thread block following the binary tree topology. In these regards, Dotsenko \textit{et al.} \cite{dotsenko2008fast} proposed a matrix-based method, \hl{which shares the same rationale with the approach proposed in \cite{blelloch1989scans}}. This method organizes the data block as a matrix and works in three steps: ($i$) locally scans each row (one row per thread with a sequential scan) and then stores the reduction per row into an auxiliary array; ($ii$) scans the auxiliary array; and ($iii$) updates the local scan per row using the corresponding element in the auxiliary array.

This matrix-based method was further optimized in \cite{yan2013streamscan} by avoiding the thread-block-level synchronization exerted between data loads from global memory and actual computation. For matrix-based methods, the number of $\oplus$ operations is subject to matrix dimensions and specific computation procedures, in other words, implementation dependent (e.g. both \cite{dotsenko2008fast}  and \cite{yan2013streamscan} assign one thread to compute one row in sequential \hl{as stated in their respective papers}). Nonetheless, we can still calculate that these methods require roughly $2n$ $\oplus$ operations as well as $2n$ data loads and $n$ stores (significantly fewer than the work-efficient method in theory). Fig.~\ref{fig:matrix_scan} shows the computational pattern of the matrix-based method.
\begin{figure}[!h]
\centering
\includegraphics[width=0.95\linewidth]{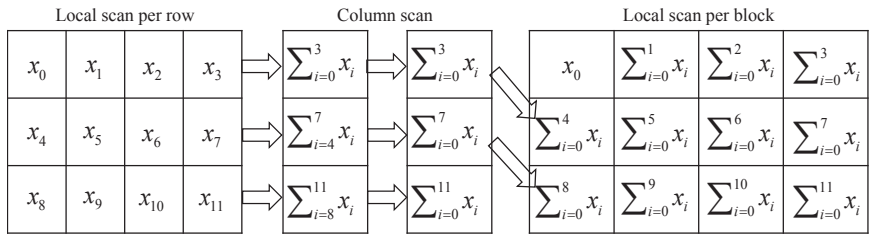}
\caption{Computational pattern of the matrix-based parallel scan}
\label{fig:matrix_scan}
\end{figure}
\section{CUDA-enabled GPU Architecture}
A CUDA-enabled GPU is a shared-memory processor with many scalar CUDA cores that are organized into a set of multi-threaded SMs. The GPU architecture has evolved through four generations: Tesla, Fermi, Kepler \cite{kepler} and the latest Maxwell \cite{maxwell}. In the following, we will briefly describe some key features of the Kepler and Maxwell architectures as their compute capability is $\geq 3.0$.

For Kepler, each SM consists of 192 CUDA cores, all of which share 64 KB configurable on-chip memory serving as shared memory and L1 cache. The per-SM L1 cache only caches data for its corresponding SM. In addition to L1 caches, Kepler offers a unified L2 cache that provides caching service across the chip (see Fig.~\ref{fig:gpu_memory}(a)). Global memory is not cached by L1 cache on Kepler, whereas read-only global memory can be optionally cached by the 48 KB read-only cache by using the \textit{const \_\_restrict} keyword. Writable global memory can only be cached by L2 cache.

Maxwell enhances Kepler and targets to improve energy efficiency without requiring significant increases in available parallelism per SM from the application. In contrast with Kepler, Maxwell introduces some new architectural changes/enhancements such as unified L1/texture cache, opt-in L1 caching of global loads, and larger shared memory to improve occupancy. Fig. \ref{fig:gpu_memory}(b) shows the memory/cache hierarchies for Maxwell. Although herein we have merely used a Kepler-based GPU, our algorithm presented here is supposed to work on Maxwell-based GPUs, because of the consistent L1/L2 cache mechanism between Maxwell and Kepler as well as the backward compatibility of instruction set architecture (ISA).
\begin{figure}[!h]
\centering
\begin{minipage}[b]{0.49\linewidth}
\subfigure[]{\label{gpu:kepler}\includegraphics[width=0.8\linewidth]{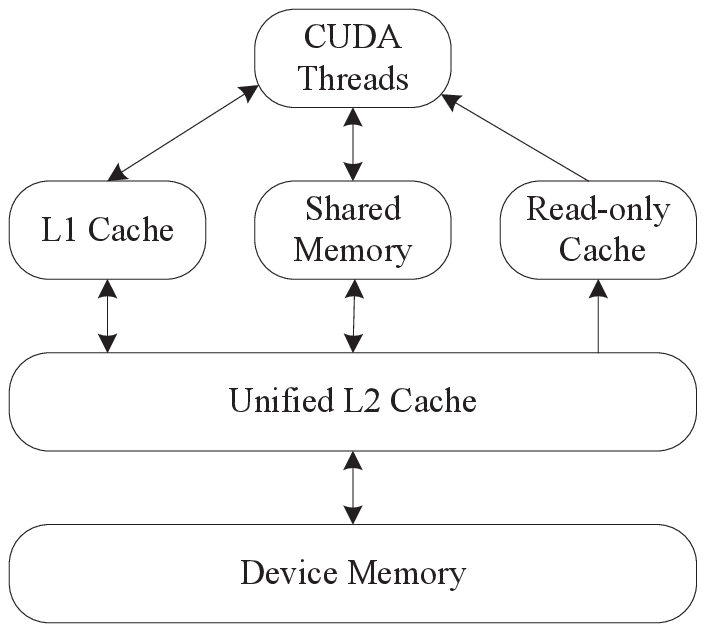}}
\end{minipage}
\begin{minipage}[b]{0.49\linewidth}
\subfigure[]{\label{gpu:maxwell} \includegraphics[width=0.8\linewidth]{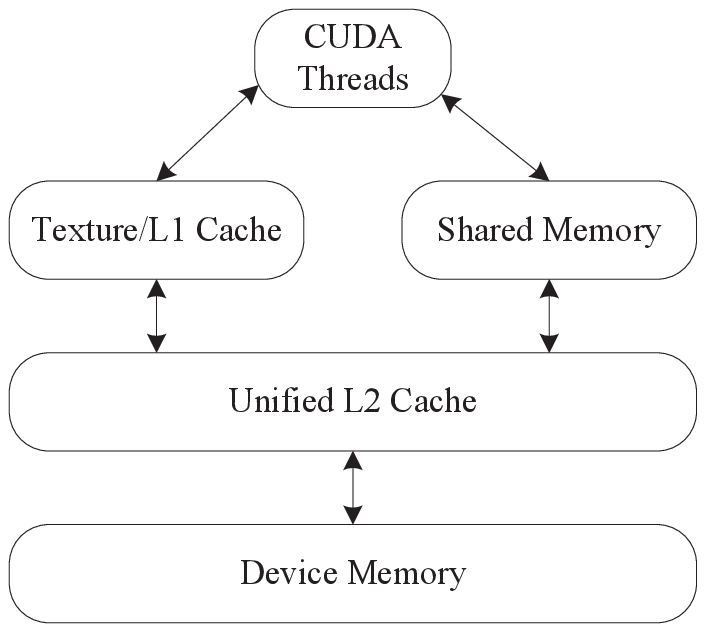}}
\end{minipage}

\caption{Memory and cache hierarchy for (a) Kepler and (b) Maxwell}
\label{fig:gpu_memory}
\end{figure}

\section{Parallelized Implementation Using CUDA}
Our algorithm works by decomposing the input array $x$ into a set of $M$ data blocks of length $L$ elements each, and then distributing them over a fixed number of $B$ (\mbox{$0\leq B \leq M$}) thread blocks in a round-robin fashion. For the array decomposition, $M$ is calculated as $\lceil\frac{N}{L}\rceil$ and $N$ is assumed to be a multiple of $L$ for simplicity. For our cyclic distribution, thread block $i$ (\mbox{$0\leq i < B$}) always receives the prefix reduction $R_{lt}$ computed by thread block $(i-1+B)\% B$, and then sends a new prefix reduction $R_{rt}$ computed by itself to thread block $(i+1)\% B$. A special case is the first data block, whose computation is independent of any other block  and which is assigned to thread block $0$ according to our distribution policy. For this special block, thread block $0$ does not need to wait for any data, and only needs to transfer its $R_{rt}$ to thread block $1$.

Basically, our cyclic distribution is based on the same underlying concept with the aforementioned chaining method \cite{yan2013streamscan}, both of which are inspired by the serial dependency between blocks. Nevertheless, they have significant differences. The chaining method launches $M$ thread blocks with one-to-one correspondence between data blocks and thread blocks. In contrast, we launch a fixed number of thread blocks and let each thread block process one data block in each iteration. For the chaining method, we have to use atomic operations to dynamically distribute data blocks to active thread blocks. This is because on CUDA-enabled GPUs, the execution order of thread blocks is non-deterministic and thread blocks with smaller indices are not guaranteed to be scheduled to run earlier than the ones with larger indices. If directly mapping thread block $i$ (\mbox{$0\leq i < M$}) to process data block $i$, we almost surely encounter a deadlock. On the contrary, our approach ensures deterministic correspondence between data blocks and thread blocks, and also makes sure that each thread block is always active during the computation. In this way, we not only can avoid the use of atomic operations for thread block re-indexing, thereby reducing computational overhead, but also evade the risk of deadlock. In practice, our cyclic distribution does demonstrate better performance than the chaining method.

In our implementation, each thread block processes one block at a time as mentioned above. Given a data block, the corresponding thread block performs an intra-block local scan, busy-waits the prefix reduction value $R_{lt}$ from its left neighbor, sends the newly computed prefix reduction value $R_{rt}$ to its right neighbor, and finally completes the scan operation corresponding to this block by updating its local scan with $R_{lt}$. Algorithm \ref{alg:workflow} shows the pseudocode for our CUDA-based scan kernel based on cyclic distribution. From the code, it can be seen that the pipeline used to process each data block comprises four stages, namely intra-warp local scan, intra-block local scan, inter-block communication and intra-block global scan.
\begin{algorithm}[!h]
\caption{Scan kernel with cyclic distribution}
\label{alg:workflow}
\begin{algorithmic}[1]
\fontsize{8pt}{8.05pt}\selectfont
\Procedure{scan\_kernel}{$x$, $y$, $M$, $\oplus$, $\cdots$}
    \For{$i$ = $blockIdx.x$; $i < M$; $i$ += $gridDim.x$}
    
        \LineComment{Performs intra-warp local scan (Algorithm \ref{alg:warp_local_scan})}
        \State{intra\_warp\_local\_scan($\cdots$);}
        
        \LineComment {Performs intra-block local scan (Algorithm \ref{alg:block_local_scan})}
        \State{intra\_block\_local\_scan($\cdots$);}
        
        \LineComment{Performs inter-block communication (Algorithm \ref{alg:block_comm})}
        \State{inter\_block\_comm($\cdots$);}
        
        \LineComment{Performs intra-block global scan (Algorithm \ref{alg:block_global_scan})}
        \State{intra\_block\_global\_scan($\cdots$)}
    \EndFor
\EndProcedure
\end{algorithmic}
\end{algorithm}
\subsection{Intra-warp Local Scan} 
In our implementation, intra-warp local scan employs warp shuffle functions as fundamental building blocks and performs the scan completely in registers. To enable more in-core computation, each thread is configured to load multiple elements, say $K$ elements, and stores all of them in $K$ register variables. In this way, each warp will process a total of $32K$ consecutive elements in $x$. Generally, our intra-warp local scan works \hl{by three steps, which are described as follows. In Step 1,} the warp loads $32K$ consecutive elements from $x$ in a coalesced manner. This is realized by $K$ iterations with each iteration loading $32$ elements by the warp. More specifically, letting $X$ denote the address of the first element corresponding to the warp, thread $i$ (\mbox{$0\leq i <32$}) within the warp loads the element $X_{i + 32j}$ in the $j$-th iteration of loads ($0\leq j < K$), and finally holds a set of $K$ elements: $\{X_{i}, X_{i + 32}, \ldots, X_{i + 32(K-1)}\}$. In this case, our coalesced data loading has actually re-organized the $32K$ consecutive elements in the form of a $K\times 32$ matrix. In our implementation, $K$ is pre-determined by taking into the following four factors: ($i$) the number of threads per thread block, ($ii$) the maximum number of 32-bit registers per thread block, ($iii$) the number of 32-bit registers per SM, and ($iv$) the size of the primitive data type used (e.g. a 32-bit register variable occupies one 32-bit register and a 64-bit register variable uses two registers.).
\begin{algorithm}[!h]
\caption{Intra-warp local scan with \mbox{$K=4$}}
\label{alg:warp_local_scan}
\begin{algorithmic}[1]
\fontsize{8pt}{8.05pt}\selectfont
\Procedure{intra\_warp\_local\_scan}{$X$, $\oplus$}

	\LineComment{Compute lane ID within the warp}
	\State {$ld$ = $threadIdx.x \& 31$;}
    \LineComment{Load elements}
	\State {$e_1$ = $X_{ld}$;}
    \State {$e_2$ = $X_{ld + 32}$;}
    \State {$e_3$ = $X_{ld + 64}$;}
   	\State {$e_4$ = $X_{ld + 96}$;}
    
    \LineComment{Scan each row}
    \State{\#pragma unroll}
    \For{$i$ = 1; $i \leq 32$; $i$ *= 2}
    	\State{$tmp$ = {\tt \_\_shfl\_up}($e_1$, $i$);}
        \If {$ld \geq i$}
        	$e_1 = e_1 \oplus tmp$;
        \EndIf
    \EndFor
  	\State{\#pragma unroll}
    \For{$i$ = 1; $i \leq 32$; $i$ *= 2}
    	\State{$tmp$ = {\tt \_\_shfl\_up}($e_2$, $i$);}
        \If {$ld \geq i$}
        	$e_2 = e_2 \oplus tmp$;
        \EndIf
    \EndFor
    \State{\#pragma unroll}
    \For{$i$ = 1; $i \leq 32$; $i$ *= 2}
    	\State{$tmp$ = {\tt \_\_shfl\_up}($e_3$, $i$);}
        \If {$ld \geq i$}
        	$e_3 = e_3 \oplus tmp$;
        \EndIf
    \EndFor
    \State{\#pragma unroll}
    \For{$i$ = 1; $i \leq 32$; $i$ *= 2}
    	\State{$tmp$ = {\tt \_\_shfl\_up}($e_4$, $i$);}
        \If {$ld \geq i$}
        	$e_4 = e_4 \oplus tmp$;
        \EndIf
    \EndFor
    
    \LineComment{Summing up the prefix reduction}
    \State {$e_2$ = $e_2$ $\oplus$ {\tt \_\_shlf}($e_1$, 31);}
   	\State {$e_3$ = $e_3$ $\oplus$ {\tt \_\_shlf}($e_2$, 31);}
    \State {$e_4$ = $e_4$ $\oplus$ {\tt \_\_shlf}($e_3$, 31);}
\EndProcedure
\end{algorithmic}
\end{algorithm}

\hl{In Step 2}, the warp performs a scan within each row of the aforementioned matrix, where each thread holds one element represented in register. This scan is implemented using the warp shuffle function \textit{\_\_shfl\_up()}, with no need of shared memory. As discussed in section \ref{sec:related_work}, the Hillis-Steele scan method involves the most $\oplus$ operations given the same number of elements. However, it needs to be stressed that more $\oplus$ operations do not mean worse speed, especially for single instruction, multiple data (SIMD)-based architectures on which multiple operations can be done by one SIMD vector instruction. For CUDA-enabled GPUs, all threads within a warp always execute one common instruction at a time. This implies that a warp can be treated as a SIMD vector of 32 lanes with each thread serving as a lane. In this case, for the Hillis-Steele scan, we can complete each iteration of the outer loop (refer to lines 3$\sim$7 in Fig.~\ref{fig:hs_scan}(b)) in constant time. In contrast, the work-efficient scan method could take twice more time than the Hillis-Steele method, as the former requires two rounds of sweep. Therefore, we have adopted the Hillis-Steele method for this step.

\hl{In Step 3}, after having completed the per-row scan, the $31$st thread within the warp holds the reduction value of each row. In this case, the warp can complete the scan row-by-row sequentially from top to bottom. Specifically, the second row is computed by broadcasting the reduction value of the first row to all threads within the warp and then letting each thread update its corresponding element to the second row. Consequently, the $31$st thread holds the prefix reduction ending at the second row. Likewise, to compute the third row, we broadcast the prefix reduction value ending at the second row to all threads within the warp and let each thread update its corresponding element to the third row. We repeat this procedure in ascending order of row index until all rows have been completed. For the broadcast operation, we have used the warp shuffle function \textit{\_\_shfl()}, not via shared memory either. Algorithm \ref{alg:warp_local_scan} gives the pseudocode of our intra-warp local scan with $K=4$.

\hl{In addition, in theory it is feasible as well to use the work-efficient scan method to perform intra-warp local scan directly over $32K$ elements, instead of the aforementioned three-phased method. However, the former is believed to yield worse performance than the latter. The reasons can be explained as follows. Firstly, the work-efficient method requires accessing elements via index. Hence, we have to store the $32K$ consecutive elements into a temporary array located in shared memory first, resulting in extra overhead incurred by array filling, and then finish the scan on the temporary array. Secondly, while likewise using a temporary array in shared memory, the matrix-based scan method overcomes the drawbacks of the work-efficient method, and yields superior performance to the latter in practice, as shown in \cite{dotsenko2008fast}. Thirdly, our three-phased method described above improves the matrix-based method and completes the scan in registers, instead of shared memory, as shared memory has significantly longer latency than registers. Based on these observations and explanations, we believe that our method outperforms the work-efficient method for the intra-warp local scan computation. Moreover, this conclusion can also be reflected by the inferior performance of CUDPP to our algorithm (see section \ref{sec:results_gpu}), where CUDPP implements the work-efficient method.}
\subsection{Intra-block Local Scan}
After each warp within a thread block completes its local scan, the $31$st  thread within each warp holds the reduction value of its own warp. In this case, we let the $31$st thread within each warp store its reduction value to an auxiliary array, and then perform a scan on the auxiliary array. The auxiliary array will be used to communicate data across the thread block, and thus can be allocated either in shared memory or global memory. Considering that the maximum number of warps per thread block is only $32$ for Kepler and Maxwell, an array of size $32$ elements is sufficient to meet our need. In this regard, we allocate the auxiliary array in the faster shared memory, rather than the slower global memory.

The scan operation on the auxiliary array can be directly performed in shared memory. Albeit faster than global memory, the latency of shared memory is still considerably higher than registers. Fortunately, the auxiliary array only has $32$ elements at the maximum and thus, this scan operation can be fulfilled completely in registers by a single warp. Assuming the auxiliary array has $32$ elements allocated in shared memory, our method works as follows: ($i$) lets thread $i$ within the first warp load the $i$-th element of the auxiliary array (\mbox{$0\leq i < 32$}); ($ii$) performs an intra-warp local scan within the warp using the warp shuffle function \textit{\_\_shfl\_up()}; and ($iii$) lets thread $i$ store its computed value to the $i$-th element of the array. In this way, while yielding fast speed, our approach is also able to avoid bank conflicts for shared memory accesses. Algorithm \ref{alg:block_local_scan} gives the pseudocode for the intra-block local scan directly following Algorithm \ref{alg:warp_local_scan}.
\begin{algorithm}[!h]
\caption{Intra-block local scan with \mbox{$K=4$}}
\label{alg:block_local_scan}
\begin{algorithmic}[1]
\fontsize{8pt}{8.05pt}\selectfont
\Procedure{intra\_block\_local\_scan}{$shrd$, $ld$, $e_1$, $e_2$, $e_3$, $e_4$, $\oplus$}

	\LineComment{Compute warp ID within the thread block}
	\State {$wd$ = $threadIdx.x / 32$;}
    
    \LineComment{Each warp saves its reduction}
    \If {$ld$ == 31}
    	\State{$shrd$[$wd$] = $e_4$;}
    \EndIf
    \State{{\tt \_\_syncthreads()};}
    
    \LineComment{Intra-warp local scan}
    \If {$wd$ == 0}
    	\State {$tmp$ = $shrd$[$ld$];}
        \State{\#pragma unroll}
        \For{$i$ = 1; $i \leq 32$; $i$ *= 2}
    		\State{$tmp2$ = {\tt \_\_shfl\_up}($tmp$, $i$);}
        	\If {$ld \geq i$}
        		$tmp = tmp \oplus tmp2$;
	        \EndIf
    	\EndFor
        \State{$shrd$[$ld$] = $tmp$;}
    \EndIf
    \State{{\tt \_\_syncthreads()};}
    
    \LineComment{Summing up the prefix reduction}
    \State{$tmp$ = $wd$ == 0 ? 0 :$shrd$[$wd - 1$]; }
    \State {$e_1$ = $e_1$ $\oplus$ $tmp$;}
    \State {$e_2$ = $e_2$ $\oplus$ $tmp$;}
    \State {$e_3$ = $e_3$ $\oplus$ $tmp$;}
   	\State {$e_4$ = $e_4$ $\oplus$ $tmp$;}
\EndProcedure
\end{algorithmic}
\end{algorithm}
\subsection{Lightweight Inter-block Communication}
Because of the serial dependency between data blocks, we need to communicate prefix reduction values between consecutive blocks. For the communication, except the first block, each of the other blocks is associated with a pair of variables, in \hbox{{\tt $($u, v$)$}} data type, allocated in global memory, where member variable $u$ stores the data to be communicated and $v$ indicates whether the data is ready to fetch, leading to a total of $M-1$ pairs for all blocks.

To ensure the update on a global memory variable is seen by all observing threads, we can simply use the \textit{volatile} keyword while loading data on cache-disabled GPUs. However, on cache-enabled GPUs, this approach actually does not work, because CUDA-enabled GPUs do not guarantee cache coherence across the device. In this case, atomic instructions are usually used to address this issue. Given the data type \hbox{{\tt $($u, v$)$}}, member variables $u$ and $v$ have to be updated separately, due to the limitations of atomic instructions. Typically, $u$ is updated by a regular global store and $v$ by an atomic operation. It needs to be stressed that a proper memory fence function (e.g. \textit{\_\_threadfence()}) should be used to guarantee that the global store operation on $u$ is done before the atomic operation on $v$ \cite{cudacprogramming}.

Albeit effective, atomic instructions are generally expensive and also perform redundant operations that are not needed by us. More specifically, from the PTX ISA, the syntax of an atomic instruction is \textit{atom\{.space\}.op.type~$r_d$,~[$addr$],~$r_b$}. This atomic instruction loads the original value at location $addr$ into register $r_d$, then reduces the value in $addr$ with operand $r_b$ using a specified operator, and finally stores the reduction result at location $addr$. Actually, however, our implementation only requires read and write access to each variable pair, not needing the reduction operation involved.
\begin{algorithm}[!h]
\caption{Lightweight inter-block communication}
\label{alg:block_comm}
\fontsize{8pt}{8.05pt}\selectfont
\begin{algorithmic}[1]
\Function{atomic\_read}{$addr$}
	\State {{\tt ulonglong2} $retval$;}
    \State {{\tt ulonglong2}* $ptr$;}
    
    \LineComment{Reinterpret cast}
    \State {$ptr$ = reinterpret\_cast$<${\tt ulonglong2}*$>$($addr$);}
    
    \LineComment{Load $ptr$ into $retval$ using inline assembly}
    
    \State{asm volatile ("ld.cg.v2.u64 $retval.x$ $retval.y$ $ptr$");}
    
    \LineComment{Reinterpret cast and return the value}
    
    \Return {*(reinterpret\_cast$<${\tt (u, v)}*$>$(\&$retval$);}
\EndFunction
\end{algorithmic}

\begin{algorithmic}[1]
\Procedure{atomic\_store}{$addr$, $val$}
	\State{{\tt ulonglong2}* ptr, tmp;}
    
    \LineComment{Reinterpret casts}
 	\State{$ptr$ = reinterpret\_cast$<${\tt ulonglong2}*$>$($addr$);}
  	\State{$tmp$ = *reinterpret\_cast$<${\tt ulonglong2}*$>$(\&$val$);}
    
   	\LineComment{Store $tmp$ into $ptr$ using inline assembly}
    \State{asm volatile ("st.cg.v2.u64 $ptr$ $tmp.x$ $tmp.y$");}
    
\EndProcedure
\end{algorithmic}

\begin{algorithmic}[1]
\Function{inter\_block\_comm}{$array$, $blockId$, $sum$, $\oplus$}
 \State{{\tt $($u, v$)$} $val$ = 0;}
 \If {$blockId > 0$}
 
 	\LineComment{Busy-wait until indicator $v$ becomes true}
    \State{$val$ = atomic\_read($array + blockId - 1$);}
    \While{$val.v$ == 0}
    	\State{$val$ = atomic\_read($array + blockId - 1$);}
    \EndWhile
 \EndIf
 \If{$theadIdx.x == blockDim.x - 1$}
 
 	\LineComment{Send the prefix reduction to the right neighbor}
 	\State{atomic\_store($array + blockId$, $make\_pair$($sum \oplus val.u$, 1);}
 \EndIf
 \State{{\tt \_\_syncthreads()};}
 
 \LineComment{Return prefix reduction from the left neighbor}
 
 \Return {$val.u$;}
\EndFunction
\end{algorithmic}

\end{algorithm}

Fortunately, for Kepler and Maxwell architectures, global memory data is coherent at L2 cache level, but multiple per-SM L1 caches are not coherent for global data (see the L1/L2 cache hierarchies in Fig.~\ref{fig:gpu_memory}) \cite{ptxisa43}. This means that if a thread in one SM updates a global memory variable via L1 cache (default behavior) and a second thread in another SM loads the same variable via L1 cache (also default behavior), the second thread may get stale L1 cache data normally, rather than the newly updated value by the first thread. On the other hand, if the first thread manipulates the update directly via L2 cache and the second thread reads the updated value directly via L2 cache (the stale L1 cache data will be accordingly invalidated), the second thread will surely get the correct value. Note that the update operation by the first thread must be fulfilled in a single transaction, which cannot be interrupted, in order to avoid race condition.

In our implementation, we have used the memory load instruction \textit{ld.cg} to cache loads only globally, bypassing the L1 cache, and cache only in L2 cache. Likewise, the memory store instruction \textit{st.cg} is used for global stores. These two instructions will evict any existing cache lines matching the requested address in L1 caches. Therefore, we can ensure that in the aforementioned example, the updates made by the first thread are correctly observed by the second thread. Through our evaluation, this method leads to faster speed than using atomic instructions. Algorithm \ref{alg:block_comm} gives the pseudocode for inter-block communication using Double primitive data type as an example. In this example, we reinterpret cast \hbox{{\tt $($u, v$)$}} type to {\tt ulonglong2} vector type before global stores, and reinterpret cast {\tt ulonglong2} back to \hbox{{\tt $($u, v$)$}} after global loads.

For other primitive data types such as Int32, Int64 and Float, they can be processed similarly. Note that the data type \hbox{{\tt $($u, v$)$}} must be packed and is recommended being aligned to some proper boundary, e.g. our algorithm uses this type attribute: \textit{\_\_attribute\_\_((packed, aligned(8)))}. It should be mentioned that as of writing this paper, we noticed that CUB \cite{cub} (not published in the literature yet) implements a similar idea for device-level communication but has different implementation details from ours. \hl{As for the scan implementation in CUB, we are not aware of the detailed method due to the lack of manual, but have observed that the implementation has massively used PTX assemblies, thus making the code difficult to be understood}. Nonetheless, the inclusive scan primitive implemented in CUB is still inferior to ours presented here (refer to section \ref{sec:results_gpu}). On the other hand, since no similar idea has been proposed yet in the literature, to the best of our knowledge, it is of high significance to related communities by explaining the core idea of this mechanism and detailing how it works.
\subsection{Intra-block Global Scan}
After receiving the prefix reduction from its left neighbor, each thread block preforms $\oplus$ operations between its local scan and the reduction value and completes the final global scan of the corresponding data block. Subsequently, all threads within a thread block write the global scan to the output array $y$ allocated in global memory in a coalesced way. Algorithm \ref{alg:block_global_scan} gives the pseudocode for our intra-block global scan, where $Y$ denotes the starting address of the first element corresponding to the current warp in $y$.
\begin{algorithm}[!h]
\caption{Intra-block global scan with $K=4$}
\label{alg:block_global_scan}
\fontsize{8pt}{8.05pt}\selectfont
\begin{algorithmic}[1]
\Procedure{intra\_block\_global\_scan}{$ld$, $Y$, $e_1$, $e_2$, $e_3$, $e_4$, $sum$, $\oplus$}

	\LineComment{Compute global scan}
	\State{$e_1 = e_1 \oplus sum$;}
    \State{$e_2 = e_2 \oplus sum$;}
    \State{$e_3 = e_3 \oplus sum$;}
    \State{$e_4 = e_4 \oplus sum$;}
    
    \LineComment{Output in a coalesced way}
    \State{$Y_{ld} = e_1$}
    \State{$Y_{ld+32} = e_2$;}
    \State{$Y_{ld+64} = e_3$;}
    \State{$Y_{ld+96} = e_4$;}
\EndProcedure
\end{algorithmic}
\end{algorithm}
\subsection{Data Deployment}
In our implementation, the output array $y$ must be writable and is therefore allocated in global memory. The input array $x$ is read-only and has only coalesced global loads. In this case, $x$ can either be placed in read-only global memory or texture memory. As mentioned above, in order to take advantage of the read-only cache, the keyword \textit{const \_\_restrict} must be used to hint to the compiler if $x$ is deployed in read-only global memory on Kepler. We have assessed the performance of our algorithm by storing $x$ in read-only global memory and texture memory (programmed in texture object application programming interfaces) respectively. Through our evaluations, we found that the cases using read-only global memory have slightly better performance than using texture memory. Since the read-only cache on Kepler serves read-only global memory loads, texture fetches and even both simultaneously, theoretically we would not expect to see any performance difference between read-only global loads and texture fetches. Nonetheless, since texture fetches result in slightly worse performance, we have allocated $x$ in read-only global memory. As mentioned above, our implementation supports in-place scan, i.e. $y$ can have an identical physical address to $x$. As each element of $x$ is read-only visited once and each element of $y$ is write-only updated once, the correct accesses to both $x$ and $y$ can be completely guaranteed regardless of whether cache is coherent or not across the device.

As for shared memory, there are two addressing modes on Kepler: 32-bit mode and 64-bit mode for shared memory, which can be switched at runtime. The 32-bit mode maps successive 32-bit words to successive banks, while the 64-bit mode maps successive 64-bit words to successive banks. In this regard, bank conflicts can be completely avoided by using 32-bit addressing mode for 32-bit words, and 64-bit mode for 64-bit words. However, Maxwell excludes support for 64-bit addressing mode. In this regard, we did not use the 64-bit model, although there would be bank conflicts for 64-bit words.
\section{Performance Evaluation}
\label{sec:results}
We have evaluated LightScan from the following three perspectives: ($i$) self-assessment in terms of different number of elements $N$ and primitive scalar data types, ($ii$) comparison to four leading GPU-based algorithms, i.e. CUDPP (v2.2), Thrust (in CUDA 7.0), ModernGPU (v1.1), and CUB (v1.4.1), and ($iii$) comparison to Intel TBB (integrated with Intel C++ compiler v15.0.1). For performance evaluation, we have used the commonly used {\tt add} (i.e. $\oplus$ is $+$) operator with varying $N$ (32 million, 64 million, 128 million, 256 million or 512 million) and varying primitive data types (Int32, Int64, Float or Double). To measure speed, we have used the billion elements per second (GEPS) metric calculated as \mbox{$\frac{N}{t}\times 10^{-9}$}, where $t$ is the wall-clock runtime measured in seconds. 

For the tests, three types of processing units (see Table~\ref{tab:processing_units}) $-$ an Intel E5-2650 16-core 2.0 GHz CPU, a Tesla K40c GPU, and an Intel Xeon Phi 5110P coprocessor $-$ have been used. Additionally, CUDA-based algorithms are all compiled using CUDA 7.0 in combination with GNU GCC v4.8.2, while TBB is compiled using Intel C++ compiler v15.0.1.

\hl{Note that we did not use $N$ $>$ 512 million, e.g. 1 billion, because of two reasons: ($i$) the performance change is tiny between some different values of $N$ $>512$ million, through our evaluation on the Tesla K40c and ($ii$) The Xeon Phi 5110P coprocessor has only 8 GB memory and is not able to hold 1 billion 64-bit elements in-memory, because the operating system and runtime libraries also consume some amount of memory within the Xeon Phi.}
\begin{table}[!h]
\caption{CPU and accelerators used in our evaluation}
\label{tab:processing_units}
\centering
\begin{threeparttable}
\begin{tabular}{|l||l||l||l|}
\hline
\textbf{Features} &\textbf{CPU}&	\textbf{GPU}&	\textbf{Xeon Phi}\\
\hline
Device& Intel E5-2650&	 Tesla K40c&	5110P\\ 
\hline
Cores/Threads&	$8\times2$&	$15\times192$&	$60\times4$\\
\hline
Core frequency&	2.0 GHz&	745 MHz	&		1.05 GHz\\
\hline
\multirow{2}{*}{Cache}&	\multirow{2}{*}{20 MB}&	16/32/48 KB$^\dagger$&	64 KB$^{\dagger\star}$\\
\hhline{~~--}
&	&	1.5 MB$^\ddagger$ &		30 MB$^\ddagger$\\
\hline
Memory&	128 GB&	12 GB&	8 GB\\
\hline
ECC$^*$&	On&		Off&	On\\
\hline
\end{tabular}
\begin{tablenotes}
\item $^\dagger$L1 cache size per core/SM; $^\ddagger$L2 cache size; $^\star$L1 instruction and data caches of size 32 KB each; and $^*$error correcting code.
\end{tablenotes}
\end{threeparttable}
\end{table}

\subsection{Assessment of Our Algorithm}
We have first assessed the performance of our algorithm. In LightScan, only one kernel launch is needed to complete a scan, and the number of thread blocks used by the kernel is equal to the number of SMs on the GPU. Each thread block is configured to have $T=1,024$ threads, i.e. the maximum number of threads per thread block allowed by Tesla K40c GPUs. In order to saturate the use of registers per SM, each thread within a thread block consumes 64 registers by setting $K$ to 44 and 20 for 32-bit and 64-bit data types, respectively. \hl{It needs to be stressed that for a SM on Kepler GPUs, the maximum number of resident thread blocks (i.e. 16), the maximum number of threads per thread block (i.e. 1,024), the maximum number of resident threads (i.e. 2,048) and the number of 32-bit registers (i.e. 65,536) are mutually related and also mutually restricted. In our case, as mentioned above, each thread within a thread block consumes 64 registers and a thread block is set to own 1,024 threads. This means that one thread block will consume all of the 32-bit registers on a SM during its execution, thus making the execution of all thread blocks launched to this SM be serialized. In this regard, if we set the number of thread blocks to be greater than the number SM in our kernel, the surplus thread blocks will not ever get a change to execute in our implementation, thus resulting in a deadlock.}

Table~\ref{tab:lightscan} gives the performance of LightScan on a Tesla K40c. From the table, it can be seen that the performance grows as $N$ increases for each data type. Therefore, our algorithm achieves peak performance at $N$~=~512 million, with the maximum performance of 25.5 GEPS, 12.8 GEPS, 25.7 GEPS and 13.0 GEPS for Int32, Int64, Float and Double, respectively.
\begin{table}[!h]
\centering
\caption{Performance (in GEPS) of LightScan on Tesla K40c}
\label{tab:lightscan}
\begin{tabular}{|l||l||l||l||l|}
\hline
$\textbf{N}$&	\textbf{Int32}&	\textbf{Int64}&	\textbf{Float}&	\textbf{Double}\\
\hline
32M&	24.1&	12.4&	24.5&	12.6\\
\hline
64M&	24.8&	12.6&	25.1&	12.9\\
\hline
128M&	25.3&	12.7&	25.5&	13.0\\
\hline
256M&	25.5&	12.7&	25.6&	13.0\\
\hline
512M&	25.5&	12.8&	25.7&	13.0\\
\hline
\end{tabular}
\end{table}

Moreover, by comparing the performance between 32-bit and 64-bit data types, we can observe that the performance is almost decreased by half when moving from Int32 to Int64 or from Float to Double. \hl{This observation could be explained by the following two reasons. One is that 64-bit data types double the data volume in comparison with 32-bit data types. Thus, the former would spend twice more time on accesses to $x$ and $y$. The other is that CUDA-enabled GPUs use two 32-bit registers to represent a 64-bit variable. In this case, for 64-bit data types the working set per thread block would be less than or equal to half of that for 32-bit data types.} In addition, it is observed that the performance for Float is always slightly better than for Int32 for each $N$. This could be because Float has a higher arithmetic instruction throughput than Int32, with respect to addition. More specifically, on Kepler architecture, the former has an instruction throughput of 192 operations per clock cycle per SM, whereas the throughput is only 160 for the latter \cite{cudacprogramming}.
\subsection{Comparison to GPU-based Counterparts}
\label{sec:results_gpu}
We have further compared LightScan with other leading GPU-based algorithms, including CUDPP, Thrust, ModernGPU, and CUB. These four algorithms demonstrate consistent relative rankings among themselves in terms of $N$ and data types. CUB performs best for each case and ModernGPU second best. On average, CUB yields a performance of $12.7$ GEPS, $6.5$ GEPS, $12.7$ GEPS and $6.5$ GEPS, while ModernGPU achieves a performance of $18.2$ GEPS, $8.3$ GEPS, $16.7$ GEPS and $8.4$ GEPS, for Int32, Int64, Float and Double, respectively. Correspondingly, the maximum performance is $21.8$ GEPS, $11.1$ GEPS, $21.8$ GEPS, and $11.2$ GEPS for CUB and $18.4$ GEPS, $8.3$ GEPS, $16.9$ GEPS and $8.5$ GEPS for ModernGPU.

CUDPP demonstrates superior performance to Thrust for 32-bit data types, while the latter performs better than the former for 64-bit data types. For Int32, Int64, Float and Double, CUDPP produces an average performance of $12.7$ GEPS, $6.5$ GEPS, $12.7$ GEPS and $6.5$ GEPS with the maximum performance of $13.2$ GEPS, $6.6$ GEPS, $13.2$ GEPS and $6.6$ GEPS, while Thrust gives an average performance of $10.6$ GEPS, $7.1$ GEPS, $9.5$ GEPS and $7.6$ GEPS with the maximum performance of $10.8$ GEPS, $7.2$ GEPS, $9.7$ GEPS and $7.7$ GEPS, respectively. Table \ref{tab:gpu_others} gives the performance comparison of all evaluated algorithms in terms of average and maximum performance.
\begin{table}[!h]
\centering
\caption{Performance (in GEPS) of each evaluated algorithm} 
\label{tab:gpu_others}
\begin{tabular}{|l||l||l||l||l||l|}
\hline
\textbf{Measure}&	\textbf{Algorithm}&	\textbf{Int32}&	\textbf{Int64}&	\textbf{Float}&	\textbf{Double}\\
\hline
\multirow{7}{*}{Average}
&	LightScan&	\textbf{25.0}&	\textbf{12.7}&	\textbf{25.3}&	\textbf{12.9}\\
\hhline{~-----}
&	CUDPP&	12.7&	6.5&	12.7&	6.5\\
\hhline{~-----}
&	Thrust&	10.6&	7.1&	9.5&	7.6\\
\hhline{~-----}
&	ModernGPU&	18.2&	8.3&	16.7&	8.4\\
\hhline{~-----}
&	CUB&	21.6&	11.0&	21.6&	11.2\\
\hhline{~-----}
&	TBB(CPU)&	3.0&	1.6&	3.0&	1.5\\
\hhline{~-----}
&	TBB(Xeon Phi)&	0.4&	0.4&	0.6&	0.4\\
\hline
\multirow{7}{*}{Maximum}
&	LightScan&	\textbf{25.5}&	\textbf{12.8}&	\textbf{25.7}&	\textbf{13.0}\\
\hhline{~-----}
&	CUDPP&	13.2&	6.6&	13.2&	6.6\\
\hhline{~-----}
&	Thrust&	10.8&	7.2&	9.7&	7.7\\
\hhline{~-----}
&	ModernGPU&	18.4&	8.3&	16.9&	8.5\\
\hhline{~-----}
&	CUB&	21.8&	11.1&	21.8&	11.2\\
\hhline{~-----}
&	TBB(CPU)&	3.1&	1.6&	3.1&	1.6\\
\hhline{~-----}
&	TBB(Xeon Phi)&	0.7&	0.7&	1.0&	0.6\\
\hline
\end{tabular}
\end{table}

Compared to each of the aforementioned GPU-based algorithms, LightScan yields superior performance. Firstly, LightScan demonstrates nearly constant speedups over CUDPP (running the \textit{cudppScan} subprogram) for each case. Specifically, LightScan runs $2.0$, $1.9$, $2.0$ and $2.0$ times faster on average than CUDPP for Int32, Int64, Float and Double, respectively. Secondly, compared to Thrust (running the \textit{thrust::inclusive\_scan} subprogram), LightScan demonstrates very consistent speedups with respect to each data type, where the average speedup is $2.4$, $1.8$, $2.7$ and $1.7$ for Int32, Int64, Float and Double, respectively. Thirdly, compared to ModernGPU (running the \textit{Scan} subprogram), LightScan yields an average speedup of $1.4$ for Int32, and $1.5$ for Int64, Float and Double. Fourthly, compared to CUB (running the \textit{DeviceScan::InclusiveScan} subprogram), LightScan  demonstrates roughly constant speedup for each case with speedup around $1.2$. Fig. \ref{fig:gpu_others} demonstrates the speedups over CUDPP, Thrust, ModernGPU and CUB.
\begin{figure}[!h]
\centering

\begin{minipage}[b]{0.49\linewidth}
\subfigure[]{\includegraphics[width=\linewidth]{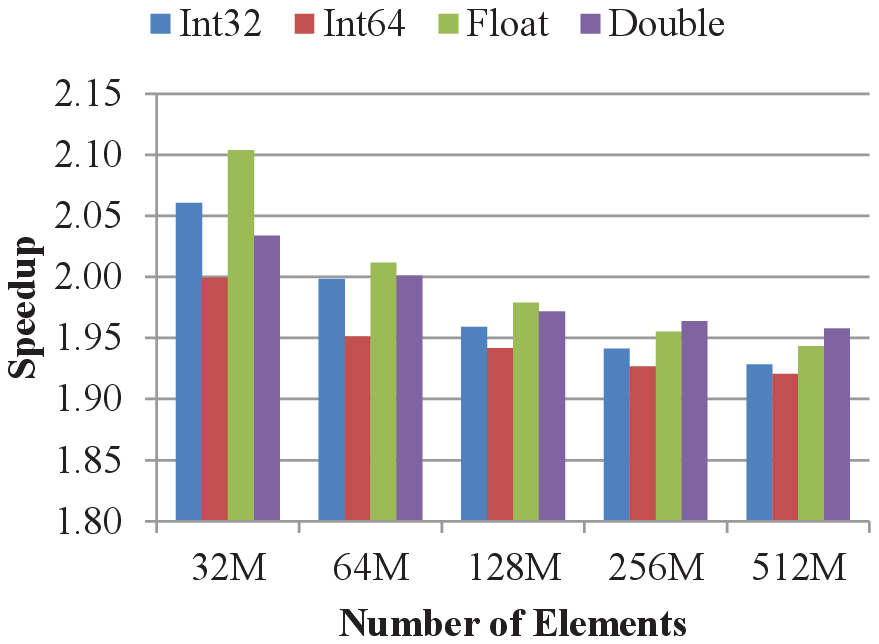}}
\end{minipage}
\begin{minipage}[b]{0.49\linewidth}
\subfigure[]{\includegraphics[width=\linewidth]{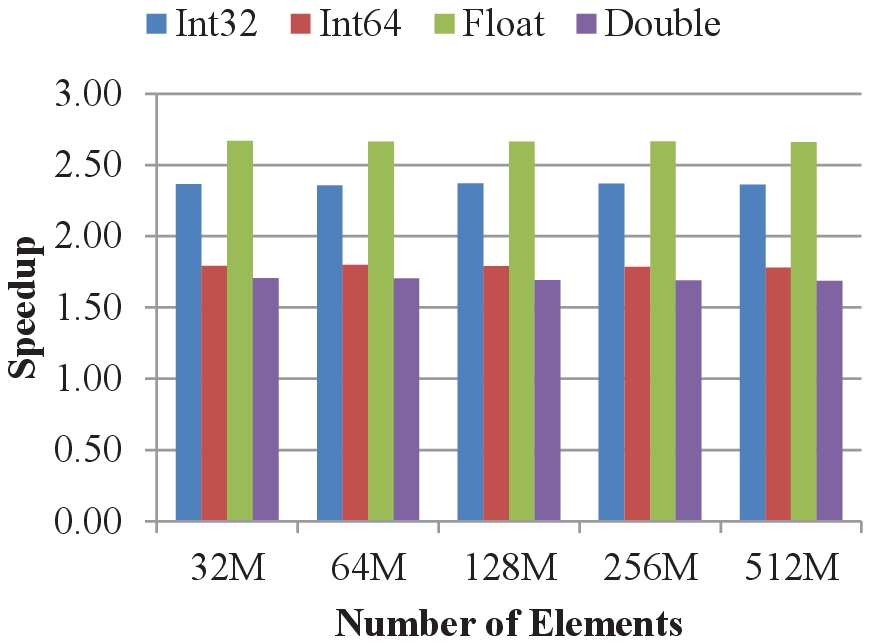}}
\end{minipage}

\begin{minipage}[b]{0.49\linewidth}
\subfigure[]{\includegraphics[width=\linewidth]{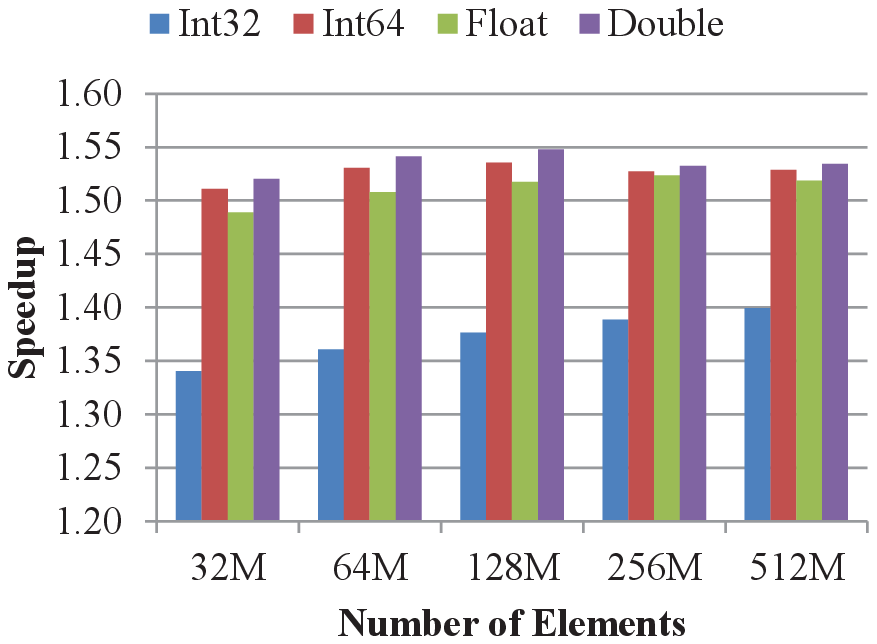}}
\end{minipage}
\begin{minipage}[b]{0.49\linewidth}
\subfigure[]{\includegraphics[width=\linewidth]{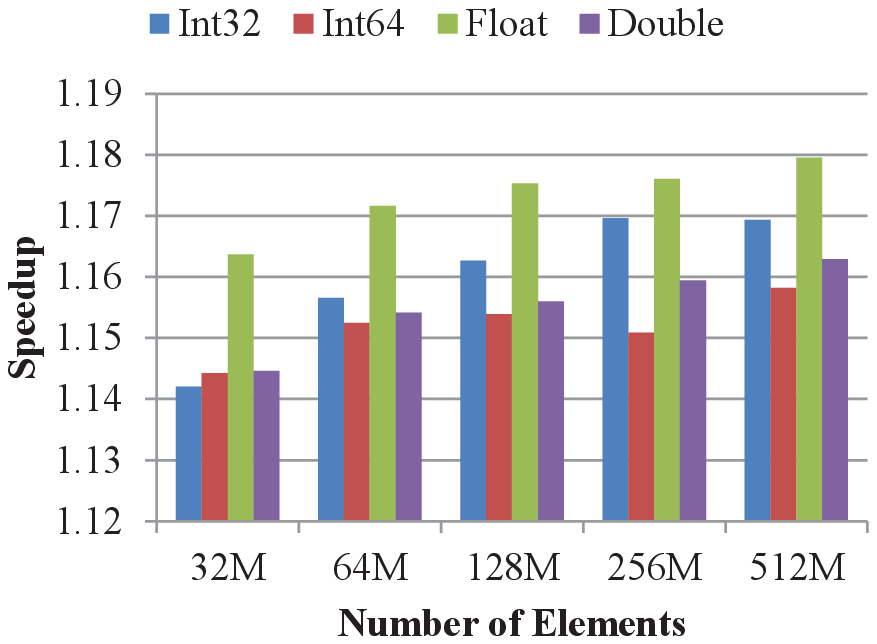}}
\end{minipage}

\caption{Performance of GPU-based algorithms: (a) CUDPP, (b) Thrust, (c) ModernGPU and (d) CUB}
\label{fig:gpu_others}
\end{figure}
\subsection{Comparison to Intel TBB}
We compared LightScan to TBB (running the \textit{paralle\_scan} subprogram), whose performance has been separately assessed on a multi-core CPU and an Intel Xeon Phi coprocessor (see Table~\ref{tab:processing_units}). On 16 CPU cores, TBB gets better performance as the number of threads $T$ increases. Thus, it reaches peak performance at $T=16$ for each case (see Fig.~\ref{fig:tbb}), with the maximum performance of $3.1$ GEPS, $1.6$ GEPS, $3.1$ GEPS and $1.6$ GEPS for Int32, Int64, Float and Double, respectively. In contrast to the GPU-based algorithms, TBB does not experience sharp performance drop on 16 CPU cores, when moving from 32-bit data types to 64-bit ones. On the Xeon Phi, TBB does not demonstrate good performance. In this test, we gain the best performance by tuning $T$ to $118$ (i.e. two threads per core) and the \textit{KMP\_AFFINITY} environment variable to \textit{balanced}. For each data type, the performance of TBB increases as $N$ becomes larger, where the peak performance is $0.7$ GEPS, $0.7$ GEPS, $1.0$ GEPS and $0.6$ GEMS for Int32, Int64, Float and Double, respectively. Table \ref{tab:gpu_others} gives the average and maximum performance of TBB on 16 CPU cores and the Xeon Phi. Note that we have used in-place scan (i.e. $y$ points to the same physical address with $x$) for the cases that device memory is not adequate in this test.

Compared to TBB, LightScan demonstrates significantly better speedups. More specifically, LightScan achieves an average speedup of $8.3$, $8.1$, $8.5$ and $8.4$ over TBB on 16 CPU cores, with the maximum speedup of $8.8$, $8.3$, $8.9$ and $8.6$, for Int32, Int64, Float and Double respectively. However, when comparing to TBB on the Xeon Phi, LightScan can achieve up to two orders-of-magnitude speedups. This reflects that the TBB library that was originally designed for Intel multi-core CPUs is not well suited to manycore Xeon Phis, although Xeon Phis theoretically have higher compute capability than multi-core CPUs. This also suggests that it is necessary to develop dedicated parallel algorithms for Xeon Phis, instead of simply applying existing parallel algorithms targeting multi-core CPUs. Fig. \ref{fig:tbb} shows our speedups over TBB on 16 CPU cores and the Xeon Phi.
\begin{figure}[!h]
\centering
\begin{minipage}[b]{0.49\linewidth}
\subfigure[]{\includegraphics[width=\linewidth]{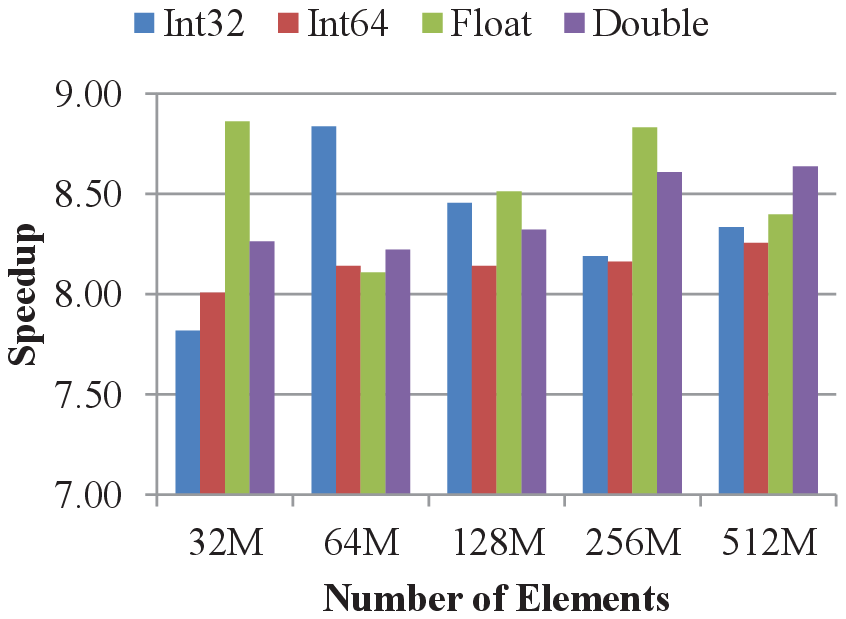}}
\end{minipage}
\begin{minipage}[b]{0.49\linewidth}
\subfigure[]{\includegraphics[width=\linewidth]{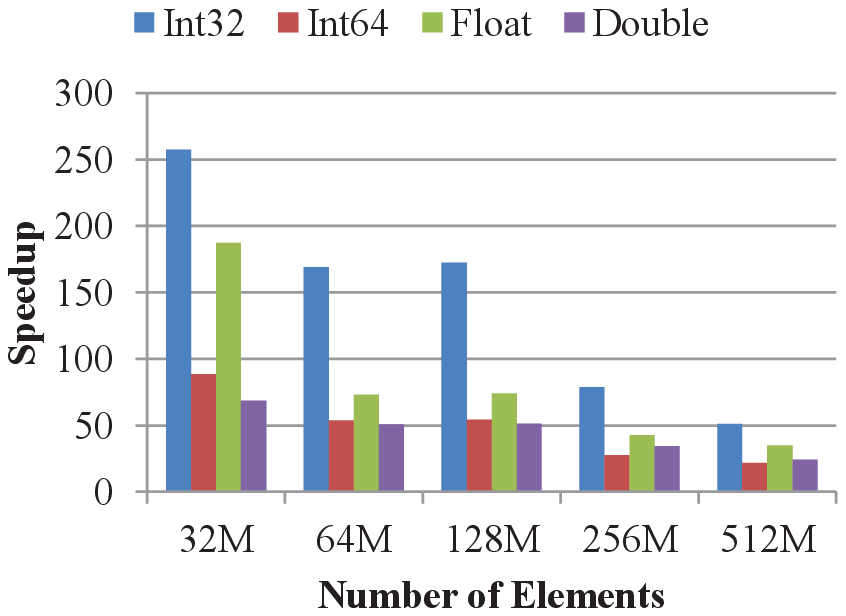}}
\end{minipage}
\caption{Performance of TBB: (a) 16 CPU cores and (b) Xeon Phi}
\label{fig:tbb}
\end{figure}
\section{Conclusion}
Scan is a frequently used primitive in various applications and its performance is critical to the overall performance of many applications to some degree. In this paper, we present LightScan, a faster CUDA-compatible parallel scan algorithm, which achieves fast speed by benefiting from warp shuffle functions and global L2 cache coherence enabled on NVIDIA GPUs with compute capability 3.0 or higher. \hl{LightScan is programmed in CUDA C++ template classes based on the generic programming paradigm, which can be instantiated with specific data types or values of the template arguments. Currently, our algorithm only supports primitive scalar data types built-in CUDA C++ language, and its source code is publicly available at \url{http://cupbb.sourceforge.net}.}

We have evaluated the performance of LightScan on a single Tesla K40c GPU by varying the number of elements and primitive data types. Performance evaluation shows that with add operator (i.e. $\oplus = +$) LightScan can scan $25.5$ GEPS, $12.8$ GEPS, $25.7$ GEPS and $13.0$ GEPS for Int32, Int64, Float and Double,\ respectively. This performance has been further compared to that of five leading algorithm, i.e. CUDPP, Thrust, ModernGPU, CUB and TBB. The first four algorithms are designed for CUDA-enabled GPUs, whereas the remaining one originally targets Intel multi-core CPUs. Compared to the four GPU-based algorithms running on the same GPU, LightScan achieves an average speedup of $2.0$, $2.1$, $1.5$ and $1.2$ over CUDPP, Thrust, ModernGPU and CUB, respectively, with the maximum speedup of $2.1$, $2.4$, $1.5$ and $1.2$ accordingly. Compared to TBB, LightScan yields an average speedup of $8.4$ and $80.8$ over the latter on 16 CPU cores and on the Xeon Phi, respectively, with the maximum speedup of $8.9$ and $257.3$.

Besides for parallel scan, the technique proposed for inter-block communication can also be used to accelerate other problems, which rely on device-level communication on GPUs, such as breadth-first search (BFS) in graph algorithms (e.g. \cite{luo2010effective}). For such cases, we would expect to achieve better performance by using our technique instead of commonly used atomic instructions. In addition, we could anticipate that our technique can be generalized for device-level peer-to-peer communications between thread blocks (or multiprocessors), by combining with the unified memory addressing in the CUDA programming model. In this regard, it would be feasible to virtualize each thread block (or each SM) as an individual process (or processor) and then allow for them to directly participate in distributed computing over GPUs.
\section*{Acknowledgment}
This research is supported in part by US National Science Foundation under IIS-1416259.


\end{document}